# Emerging disciplines based on superatoms: a perspective point of view

Jianpeng Wang,[1, 2, §] Weiyu Xie,[1, 2, §] Yang Gao,[1, 2] Dexuan Xu[1, 2] and Zhigang Wang[1, 2, *]
1 Institute of Atomic and Molecular Physics, Jilin University, Changchun 130012, China
2 Jilin Provincial Key Laboratory of Applied Atomic and Molecular Spectroscopy, Jilin University, Changchun 130012, China
§ These authors contributed equally to this work.
* E-mail: wangzg@jlu.edu.cn

Abstract: In this work, our statements are based on the progress of current research on superatomic clusters. Combining the new trend of materials and device manufacture at the atomic level, we analyzed the opportunities for the development based on the use of superatomic clusters as units of functional materials, and presented a foresight of this new branch of science with relevant studies on superatoms.

In 2016, a whole new organization for advanced science and technology was created in China [1], and even it is called China's Defense Advanced Research Projects Agency (DARPA). Despite the appropriateness of this metaphor, the reference to DARPA Atoms to Product (A2P) Program reminds us a promising trend towards the atomic-level manufacture of functional materials and devices. This ensures functional units on the atomic level to play an important role. Superatomic clusters, whose properties can be controlled at the atomic level, obviously fit into this new trend and represent an opportunity for integrating different scientific disciplines.

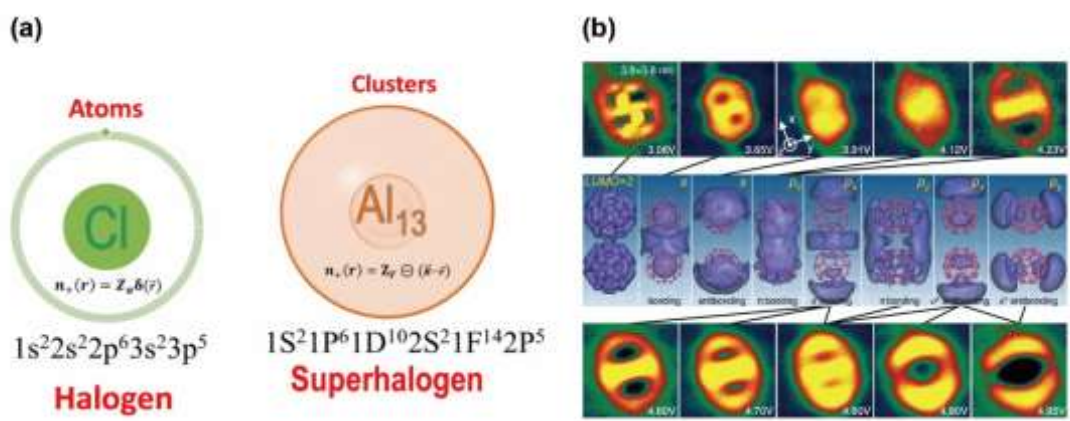

**Fig. 1** (Color online) Typical superatom structures of $Al_{13}$ and $C_{60}$. (a) The $Al_{13}$ clusters exhibit

electron shell configurations similar to that of Cl atoms [2, 3]. (b) The intermolecular interaction between $C_{60}$ superatoms shows the orbital characteristics similar to interatomic bonding [4]. Panel (a) is reprinted with permission from Ref. [2]. Copyright 2014 © American Chemical Society. Panel (b) is reprinted with permission from Ref. [4]. Copyright 2008 © American Association for the Advancement of Science.

Superatoms originated within cluster science [2-6]. Although the current understanding of their regularities needs to be deepened further, different formulations have common properties. For instance, as stable clusters, superatoms have high point group symmetry, and their molecular orbitals (i.e. electronic structures) can be very similar to that in atoms, as shown in Fig.1 [2-4]. Especially, the atom-like electronic structure properties of superatoms are not highlighted in the general cluster. This is a basic and distinct feature that can be used to distinguish the superatom from the cluster. Compared to atoms, the properties of superatoms are also influenced by nuclear movement. For example, to study the electric-dipole transition of a superatom, its wavefunciton is made up of electron wavefunction and vibration function. Hence the physical content of superatoms is higher. Consequently, the study on superatoms will not only enrich traditional atomic and molecular physics, but will also lead to the development of new research fields.

Since the atom is the smallest chemical unit and the basic constituent unit of matter, understanding atomic laws has important significance for the superatom. The study of atoms directly impacts the development of atomic physics and thus greatly promotes the birth and perfection of quantum mechanics. Thus, the solution of the electronic structure of the microsystem based on quantum mechanics which is established at the atomic level has increased the basic understanding of the material universe composed of atoms. An understanding of the atomic level plays a decisive role in chemistry, material science, life sciences, and in all applications of these basic disciplines. Being microscopic cluster structures, superatoms can exhibit some of the properties of atoms. Studying superatoms at the atomic level not only provides the object of the understanding of atomic structures, but may also be a chance to develop new disciplines.

Superatomic structures had been noted more than 30 years ago, with specific peaks of clusters appearing on the mass spectrum [7, 8]. After 2000, some landmark superatom systems have been revealed, such as $C_{60}$, $Au_{20}$, $B_{40}$ and the carbon and coinage-metal-based structures in which they were included [4, 9, 10]. Overall, we know that a wide range of elements can make up the superatom system as elementary substances, for instance, carbon, boron, and coin metals; and transition metals, actinides, and lanthanides can be endohedral complexes. Furthermore, we notice that the structure of superatom system has also entered the traditional "nano" scope. This leads to a fundamental question: can superatoms be classified as "nano"? Obviously, the development of superatomic clusters at the nanoscale cannot be separated from the progress of nanoscience and nanotechnology. Upon introduction of superatoms, however, its connotation should no longer be understood as "nano" in the general sense. These two subjects should therefore be treated separately, and there is a need to clarify the similarities and differences between nanomaterials and superatoms. One possible important aspect concerning the structural properties of the superatomic clusters is significantly different from those of materials commonly identified as nanomaterials, which are characterized by a high specific surface area. Actually, there is no clear correlation between large specific surface area and the increase or decrease in the number of atoms. However, the superatom is obviously different. Even the addition or reduction of only one atom has a significant effect on the electronic structure because of the conformational change. Thus, the properties of the superatom will change. In short, the delocalization of (valence) electrons in superatoms is highlighted at the atomic level, contrarily to what is observed in nanostructures. Consequently, the understanding, regulation, and manipulation of superatoms should occur at the atomic level.

By using the superatom as a unit, we can achieve atom-based functional features while breaking the restrictions of atom-based materials in order to achieve high performance or even revolutionary applications. The reason can be explained in this way. We know that the processes of digesting will not breakdown the intake into atoms but into amino acids, water, carbon dioxide, and other small molecules. Hence, the minimum unit of substance is the atom, but the minimum functional unit could be bigger. Superatoms could partially substitute atoms as functional units because they potentially possess atomic-like functions. Furthermore, there exist only 90 kinds of naturally stable elements, but one could say that the abundance of superatom is almost endless. Thus, it is possible to find superatoms that satisfy specific needs, to some extent surpassing the possibilities given by nature.

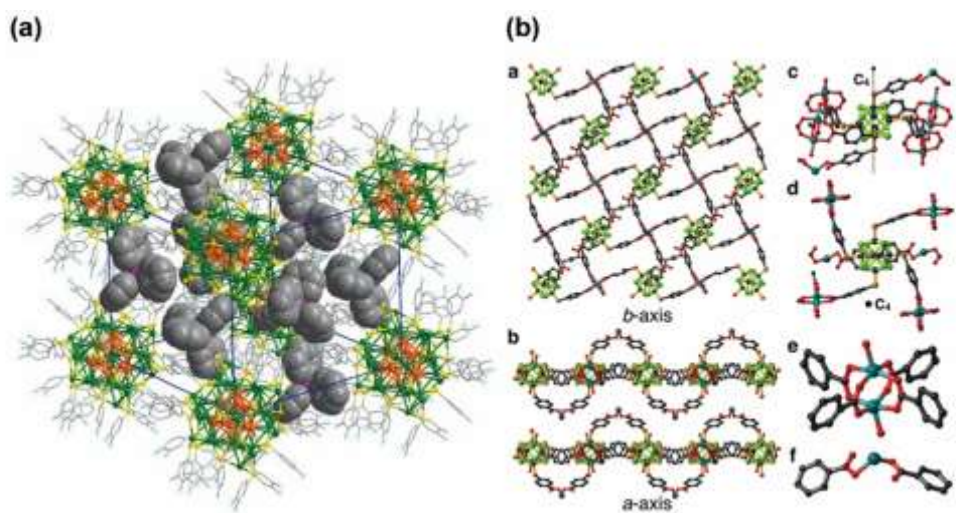

**Fig. 2** (Color online) Representative structural materials based on superatom assembly. (a) three-dimensional assembly [11]. (b) two-dimensional assembly [12]. Panel (a) is reprinted with permission from Ref. [11]. Copyright 2013 © Springer Nature. Panel (b) is reprinted with permission from Ref. [12]. Copyright 2017 © American Chemical Society.

How is a superatom made to play the role of a functional unit, except for building functional devices that consist of a single superatomic cluster? It is an important way for realizing the functionalization by assembling them from superatomic individuals to form bulk materials with certain functions, as shown in Fig.2 [11, 12]. The assembly of bulk material maintaining superatomic properties requires the support of different fields of science, including physics, chemistry, and material science. For further progress, the application of superatomic functionalization and even industrialization will need to be associated with natural science, applied science, technology, and so forth. From this point, we can see the further development of different disciplines. Even more cross fusion is needed for addressing the needs of developments in superatoms.

To illustrate the need for the development of interdisciplinary research on the subject of superatoms, we can still use atoms to do the analogy. As everyone knows, the understanding of atomic structure and spectroscopy follows quantum mechanics. For superatoms, the delocalized valence electronic orbitals could show atom-like electron shell arrangement characteristics, making it possible to describe the tools used for atoms previously. However, we know that isolated atoms possess a three-dimensional spherical symmetry, whereas it is different in superatoms, which can also exhibit a two-dimensional planar symmetry [13]. This makes the superatom different from natural atoms in terms of electron transitions. Moreover, the configuration of a superatom is based on the positional relationship between atoms, thus the nuclear vibration of these atoms is also involved in the radiating transition. In order to understand and apply the basic nature of superatoms, we must establish a new structural and spectroscopic theory to account for superatoms, in contrast to the atomic structure and spectroscopy that is commonly known. But this is faced with the problem of isolated superatomic individuals. It can be seen that despite the developments in superatoms in more than 30 years, we still have a long way to go to grasp its physical laws and even to achieve its functional application. This shows that despite the long-standing research on the subject, progress still needs to be made to achieve the functional application of superatoms.

The exploration of the superatom and even attempts for an application are still at the primary stage of development, inevitably requiring a large amount of studies. Meanwhile, the new trend of manufacturing materials and devices at the atomic level provides a key opportunity for the development of superatoms with potential as new functional elements. As long as we clarify the relationship between relevant applications and basic research, parallel development is not contradictory. Because of the necessary guidance, we can promote the clear purpose of superatomic research and a more rational allocation of related resources. We can therefore find a route that is conducive to the healthy development of superatomic research. The consequent emergence of disciplines related to superatoms can be expected.

**Conflict of interest**

The authors declare that they have no conflict of interest.

**Acknowledgment**

This work was supported by the National Natural Science Foundation of China (11674123 and 11374004) and the Science and Technology Development Program of Jilin Province of China (20150519021JH). Z. W. also acknowledges the Fok Ying Tung Education Foundation (142001) and the High Performance Computing Center of Jilin University.